# STRAIN ENGINEERING OF FERROMAGNETIC - GRAPHENE - FERROELECTRIC NANOSTRUCTURES

Eugene A. Eliseev[1], Anna N. Morozovska[2*], and Maksym V. Strikha [3,4],

1 - Institute for Problems of Materials Science, National Academy of Sciences of Ukraine,
Krjijanovskogo 3, 03142 Kyiv, Ukraine

2 - Institute of Physics, National Academy of Sciences of Ukraine,
Pr. Nauky 46, 03028 Kyiv, Ukraine,

3 - Taras Shevchenko Kyiv National University, Faculty of Radiophysics, Electronics and Computer Systems, Pr. Akademika Hlushkova 4g, 03022 Kyiv, Ukraine,

4 - V.Lashkariov Institute of Semiconductor Physics, National Academy of Sciences of Ukraine, Pr. Nauky 41, 03028 Kyiv, Ukraine

## ABSTRACT

We calculated a spin-polarized conductance in the almost unexplored nanostructure "high temperature ferromagnetic insulator/ graphene/ ferroelectric film" with a special attention to the impact of electric polarization rotation in a strained multiaxial ferroelectric film. The rotation and value of polarization vector are controlled by a misfit strain. We proposed a phenomenological model, which takes into account the shift of the Dirac point due to the proximity of ferromagnetic insulator and uses the Landauer formula for the conductivity of the graphene channel. We derived analytical expressions, which show that the strain-dependent ferroelectric polarization governs the concentration of two-dimensional charge carriers and Fermi level in graphene in a self-consistent way. We demonstrate the realistic opportunity to control the spin-polarized conductance of graphene by a misfit strain ("strain engineering") at room and higher temperatures in the nanostructures $CoFeO_4$/graphene/PZT and $Y_3Fe_5O_{12}$/graphene/PZT. Obtained results open the possibilities for the applications of ferromagnetic/graphene/ferroelectric nanostructures as non-volatile spin filters and spin valves.

---

* Corresponding author: anna.n.morozovska@gmail.com

# I. INTRODUCTION

Shortly after a field transistor with a graphene channel on a dielectric substrate was created in 2004 for the first time [1], multiple attempts have been made to use the unique properties of the new 2d-material in spintronics. At first graphene was proposed to be used as a non-magnetic spacer connecting two ferromagnetic contacts of the spin valve [2]. It has been experimentally shown that, due to the small spin-orbital interaction in graphene, the spin-relaxation length of a spin-polarized current at room temperature can be of 2 μm order [3].

However, at the same time, it was concluded that graphene is poorly attractive for spintronics, since the magnetoresistance of the valve is small due to the small number of conductance modes corresponding to the graphene channel Fermi energy in comparison with the analogous number of modes in ferromagnetic contacts [3]. Despite of pessimistic expectations effective spin valves with a graphene "spacer" and cobalt contacts have been created soon [4]. Since then, intensive efforts have been made to improve such valves. In particular, quite recently, the spin valve with cobalt contacts and 6 μm graphene channel have been created [5], at that spin polarization of the injection contact can be controlled by the bias and gate voltages [6].

In parallel, an effective device has been proposed, which is either a spin valve or a spin filter, and no longer uses graphene as a nonmagnetic spacer, but as an active ferromagnetic element [7]. To realize this, an insulator ferromagnetic EuO is imposed on the part of the graphene channel, which results in the strong spin polarization of the π-orbitals of graphene. As a result, the splitting of the graphene band states into the subbands with the orientation of the spin values "up" and "down" occurs, and also EuO induces the energy gap between these bands [8, 9]. The transition between the states of the filter and the valve in [10] was induced by voltage at the lower gate.

It was demonstrated how the ferroelectric substrate with an out-of-plane spontaneous polarization can be used for the doping of a graphene conductive channel by a significant number of carriers without the traditional application of the gate voltage to the dielectric substrate [11, 12].

Recently Kurchak et al. [13] have shown that the nanostructure $EuO/graphene/BaTiO_3$ can operate as a spin valve at temperatures well below the Curie temperature of EuO. However the work [13] is a model one and does not take into consideration several physical effects, which are critically important for the correct understanding of the valve (or filter) operation and its real applications in spintronics.

The first effect is the spontaneous polarization rotation in multiaxial ferroelectrics, like $BaTiO_3$ or $Pb(Zr,Ti)O_3$, induced by a misfit strain [14, 15] originated from the lattices mismatch of the ferroelectric film and its substrate. Actually, only out-of-plane component of the spontaneous polarization can induce significant number of carriers in graphene, while the in-plane component does not affect the carrier density at all.

The second effect is the big (more or about 1 eV) spin-independent shift of the Dirac point induced by the proximity of ferromagnetic insulator [8-10]. The compensation of the shift demanded a mandatory presence of the top gate [13] and consequently precludes simulation of the "true" non-volatile valve. However, the top gate was not the biggest disadvantage of the nanostructure $EuO$/graphene/$BaTiO_3$, but the strict requirements of a very small spontaneous polarization (less than $mC/m^2$) to move the Fermi level of graphene inside the small band gap induced by a ferromagnet. Such small polarizations possibly exist only in the immediate vicinity of the strain- or size-induced phase transitions in a thin ferroelectric film; and their values are very hard to control by e.g. tiny changes of misfit strains.

The third effect is the low Curie temperature of EuO, $T_C = 77$ K, that prevents the spin-polarized filtration at room and elevated temperatures. The obstacle can be readily overcome by using high temperature ferromagnetic insulators, such as $CoFeO_4$ ($T_C = 793$ K) and $Y_3Fe_5O_{12}$ ($T_C = 550$ K), for which the proximity effect arising in the band structure of $CoFeO_4$/graphene and $Y_3Fe_5O_{12}$/graphene are calculated from the first principles in Refs.[8-10].

In this work we consider the spin-dependent conductance in a very poorly studied nanostructure "high temperature ferromagnetic insulator/ graphene channel/ strained ferroelectric film" with a special attention to the effects caused by the rotation of electric polarization in multiaxial ferroelectrics due to the misfit strain. Our analytical calculations included a spin-independent shift of the Dirac point induced by the proximity of ferromagnetic insulator. Using a self-consistent approach, we analyzed how a strain-dependent ferroelectric polarization influences the concentration of two-dimensional charge carriers and Fermi level in graphene.

The proposed phenomenological model is presented in **Section II**, where we describe the considered geometry, discuss the effective Hamiltonian of the graphene in the proximity of ferromagnetic insulator in comparison with an isolated graphene, analyze the Landauer formula for the conductivity of the graphene channel, and study the influence of the ferroelectric polarization on two-dimensional (**2D**) carriers in graphene. The influence of misfit strain and temperature on the Fermi energy is discussed in **Section III**. The possibility to control the spin-dependent conductance of ferromagnetic-graphene-ferroelectric nanostructure by a misfit strain ("strain engineering") is discussed in **Section IV**. **Section V** is a brief summary.

## 2. PHENOMENOLOGICAL MODEL

### A. The geometry of the problem

Graphene single-layer of length $L$ and width $W$ is placed between a single-domain ferromagnetic insulator and a polarized ferroelectric film of thickness $h$ (see **Fig. 1a**). The gap between the ferromagnet and graphene is absent in order to provide maximal proximity effect. In contrast, we regard that an ultra-thin dielectric gap of thickness $d$ exists between the ferroelectric and graphene. The gap allows to avoid the proximity effect of graphene and ferroelectric atomic wave functions.

The out-of-plane spontaneous polarization $+P_3$ corresponds to the positive bound charge at the graphene-ferroelectric interface. The value of $P_3$ determines the variation of the free carrier density in the graphene channel, namely the proportionality $\delta\sigma \sim P_3$ exists. In fact, $|\delta\sigma| < |P_3|$, due to the electric field drop in the gap and incomplete screening of the bound charge by a single-layer graphene with a very small (but finite) screening length.

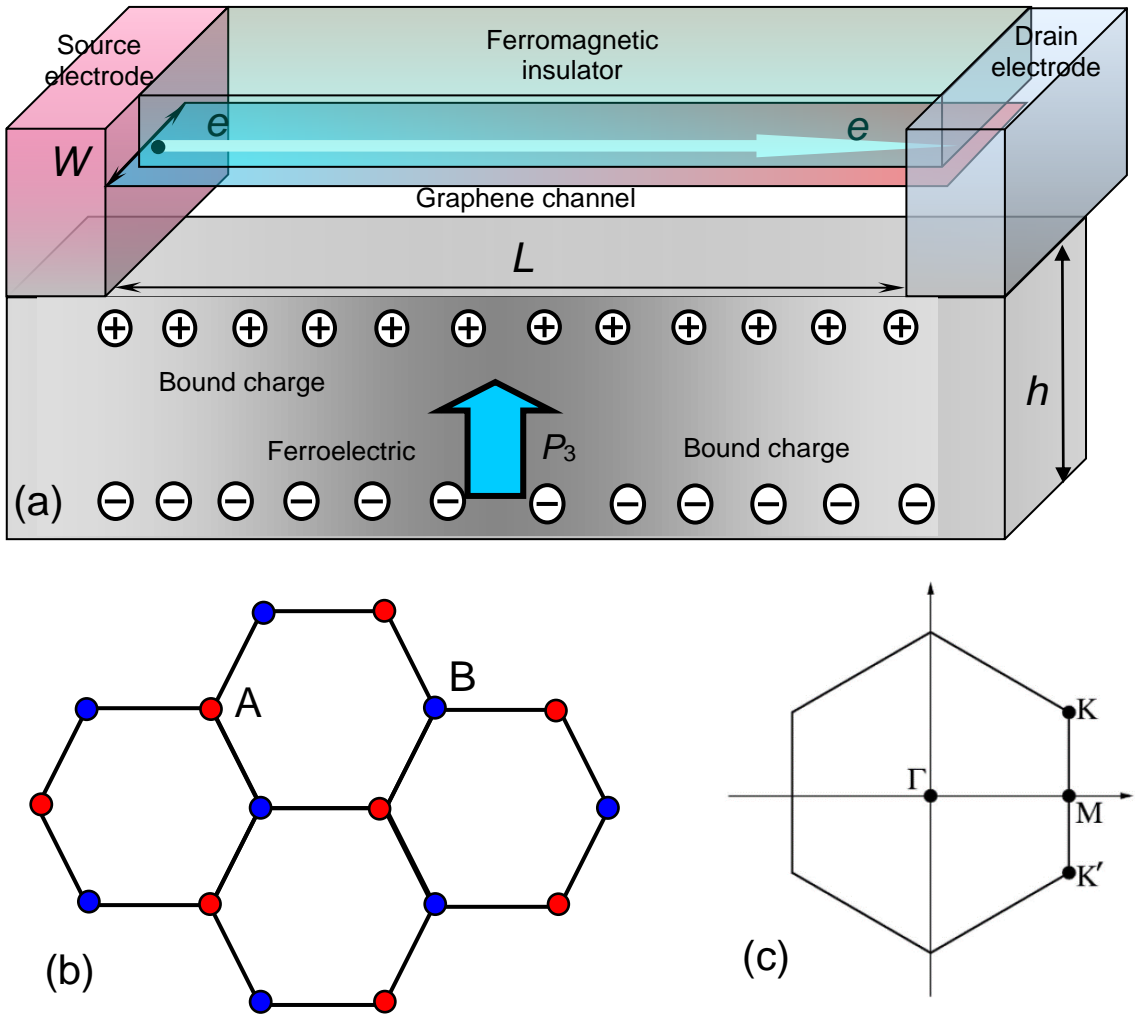

**Fig.1. (a)** Graphene single-layer of length $L$ and width $W$ is placed between a single-domain ferromagnetic insulator and a polarized ferroelectric film of thickness $h$. Adapted from Ref. [13]. **(b)** Bravais lattice with primitive vectors $a_1$ and $a_2$ (lattice constant 0.142nm); marked sublattices A and B; **(c)** the first Brillouin zone with marked corners $K$ and $K'$ of different symmetry. Adapted from Ref. [16].

## B. Effective Hamiltonian of an isolated graphene

A graphene honeycomb structure comprises two equivalent carbon sublattices A and B with charge carriers described by massless Dirac fermions. There are the two Dirac points $K$ and $K'$ at the corners of the graphene Brillouin zone (see **Fig. 1b** and **1c**). In the vicinity of these two points, **k = K +q** and **k = K' +q,** the electronic structure of graphene is characterized by a linear dispersion relation with the Dirac point separating the valence and conduction bands with a zero-band gap as follows:

$$H(\boldsymbol{q}) = \hbar v_F \boldsymbol{q}\hat{\boldsymbol{\sigma}}, \quad (1a)$$

Here $\hbar = 6.583 \times 10^{-16}$ eV·s, **q** is the wavevector, $v_F = 10^6$ m/s represents the Fermi velocity, that does not depend on the energy or momentum, and $\hat{\boldsymbol{\sigma}}$ are Pauli matrices. According to Eq.(1a), an isolated single-layer graphene is a 2D gapless semiconductor with a linear band spectrum near the Dirac points:

$$E^{\pm}(q) = \pm\hbar v_F q, \quad (1b)$$

where $q = \sqrt{q_x^2 + q_y^2}$ is the momentum measured relative to the Dirac point and, and the signs "+" and "-" correspond to the conduction and valence bands, respectively. The gapless Dirac cones at K and $K'$ are protected by time reversal and inversion symmetry. Since Dirac points are separated in the Brillouin zone, small perturbations cannot lift this valley degeneracy [9].

## C. Effective Hamiltonian of a graphene in the proximity of ferromagnetic insulator

In accordance with ab initio calculations [8 - 10] the gapless spectra (1) of the graphene channel section undergoes specific modifications in the proximity of ferromagnetic insulator. In particular, graphene sublattices A and B feel different chemical environment in the proximity of a ferromagnetic insulator, which leads to the inversion symmetry breaking between $K$ and $K'$ points and induces the spin-dependent band gap opening. The proximity effect can be modelled by the following effective Hamiltonian [10]:

$$\widehat{H}_s^{\pm}(q) = \hat{\sigma}_0(D_0 + D_s) + \hbar v_s \hat{\boldsymbol{\sigma}} \boldsymbol{q} \pm \Delta_s \hat{\sigma}_z, \quad (2a)$$

where the subscript $s = \uparrow, \downarrow$ designates the two values of the "up" and "down" spin projections; $\hat{\sigma}_0$ is a unit matrix, $D_0$ is the spin-independent shift of the Dirac point induced by the exchange coupling between the graphene sublattices and magnetic moment of magnetic atoms, $D_\uparrow$ and $D_\downarrow$ are "up" and "down" shifts of the Dirac point, spin-dependent Fermi velocities are $v_\uparrow = 1.15 \cdot 10^6$ m/s and $v_\downarrow = 1.40 \cdot 10^6$ m/s. The spin-orbital coupling term is expressed via the Pauli matrix $\hat{\sigma}_z$. The splitting energies $\Delta_\uparrow$ and $\Delta_\downarrow$ determine the spin-dependent gap opening $\frac{\Delta_\uparrow+\Delta_\downarrow}{2}$, and the "mass" term $\frac{\Delta_\uparrow-\Delta_\downarrow}{2}$ introduced in Ref.[9].

Parameters of the effective Hamiltonian Eq.(2a) depends on the ferromagnet material, graphene-ferromagnet lattices orientation, as well as on the thickness of ferromagnetic insulator [9]. The band structure of graphene in the proximity of a ferromagnetic insulators EuO, CoFe$_2$O$_4$, EuS and Y$_3$Fe$_5$O$_{12}$ were calculated from the first principles [8 - 10]. The results [8 - 10] have been interpolated by the analytical dependence for the energy levels [10]:

$$E_s^\pm(q) = D_0 + D_s \pm \sqrt{(\hbar v_s q)^2 + (\Delta_s/2)^2}. \tag{2b}$$

The characteristic energies of the band edges arising in the band spectrum $E_s^\pm(q)$, at $q=0$ are $E_\uparrow^\pm(0) = D_0 + D_\uparrow \pm \frac{\Delta_\uparrow}{2}$ and $E_\downarrow^\pm(0) = D_0 + D_\downarrow \pm \frac{\Delta_\downarrow}{2}$. Notably that the shift $D_0$ was not included in the calculations [13], and its compensation demanded mandatory presence of the top gate and consequently precluded to model the "true" non-volatile valve.

The dependence of the energy levels (2) on the wave vector **q** for CoFeO$_4$/graphene and Y$_3$Fe$_5$O$_{12}$/graphene nanostructures are shown in **Fig. 2a** and **2b**, respectively. Blue and red curves correspond to spin down and spin up states, respectively. Parameters in Eqs.(2) for the pairs CoFeO$_4$/graphene and Y$_3$Fe$_5$O$_{12}$/graphene are listed in **Table I**. They were taken from Song et al. [10] and Hallal et al. [9] *ab initio* results, as described in **Appendix A1**.

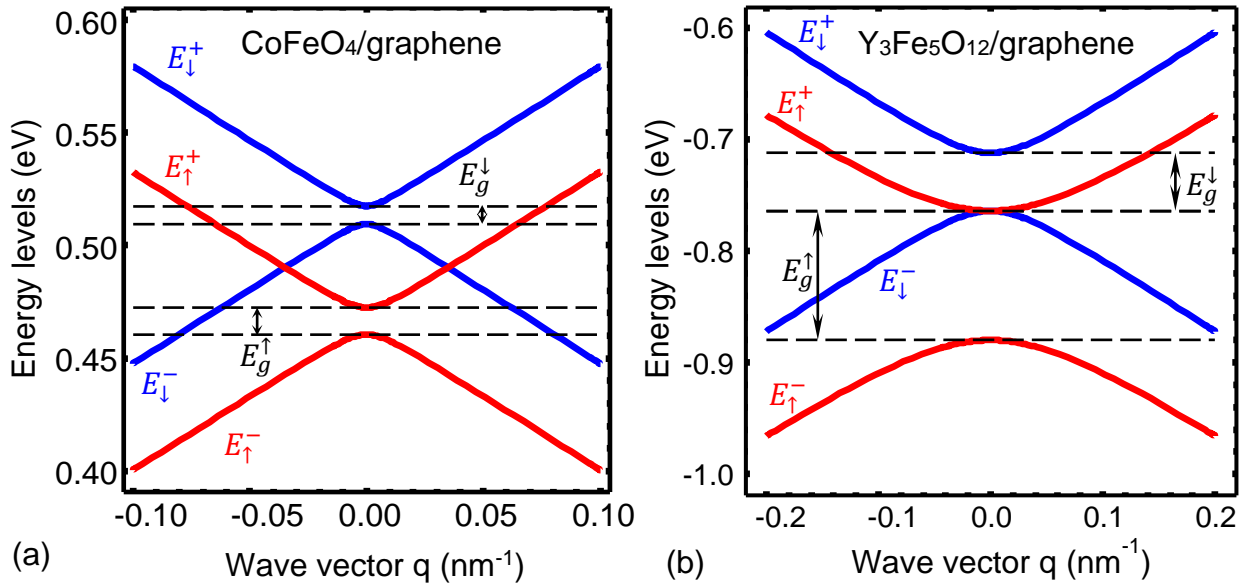

**FIGURE 2**. Energy levels (2b) dependence on the wave vector q for CoFeO$_4$/graphene **(a)** and Y$_3$Fe$_5$O$_{12}$/graphene **(b)** nanostructures. Blue and red curves correspond to spin "down" and "up" states, respectively. Parameters for the pairs CoFeO$_4$/graphene and Y$_3$Fe$_5$O$_{12}$/graphene are listed in **Table I**.

**Table I.** Parameters for Eqs.(2) taken from Refs.[8-10], as described in **Appendix A**.

| Material | $D_0$ (eV) | $D_\uparrow$ (meV) | $D_\downarrow$ (meV) | $\Delta_\uparrow$ (meV) | $\Delta_\downarrow$ (meV) | $T_C$ (K) | Ref. |
|---|---|---|---|---|---|---|---|
| EuO/Gr * | −1.36(7) | 31 (33) | − 31 (-33) | 134 | 98 | 77 | [8, 10] |
| $Y_3Fe_5O_{12}$/Gr | − 0.78 | −42 | +42 | 116 | 52 | 550 | [9] |
| $CoFeO_4$/Gr | +0.49 | − 23.5 | +23.5 | 12 | 8 | 793 | [9] |

"Gr" – graphene

One can see from **Fig. 2** and **Table I**, that the spin-dependent splitting of Landau levels, $E_g^s = E_s^+ - E_s^- \equiv \Delta_s$ is much higher for $Y_3Fe_5O_{12}$/graphene ($E_g^\uparrow = 116$ meV, $E_g^\downarrow = 52$ meV ) in comparison with $CoFeO_4$/graphene ($E_g^\uparrow = 12$ meV, $E_g^\downarrow = 8$ meV). The effective masses $m_s = \mp \frac{\Delta_s}{2v_s^2}$ of graphene carriers near subband edges are much smaller for $CoFeO_4$ in comparison with $Y_3Fe_5O_{12}$.

### D. Landauer formula for the conductivity of the graphene channel

The full conductivity of the graphene channel, taking into account the double degeneration of graphene at points $K, K'$, will be described by the modified Landauer formula [17, 18]:

$$G = \sum_s G_s, \qquad G_s = G_0 M_s(E_F) T_s(E_F). \qquad (3)$$

Here the summation is over both spin values $s = \uparrow, \downarrow$. The conductance $G_0 = \frac{e^2}{2\pi h}$ is the reverse Klitzing constant. $M_s(E_F)$ is the number of conductance modes, $T_s(E_F)$ is the transmission coefficient of the "ferromagnetic" section of length $l$, equal to the probability that the electron will pass it without scattering.

The Fermi energy $E_F$ defined by the 2D-concentrations of electrons $n$ and holes $p$ as:

$$n(E_F) = \int_{-\infty}^{+\infty} g_G^+(E) f(E - E_F) dE, \qquad p(E_F) = \int_{-\infty}^{+\infty} g_G^-(E) f(-E + E_F) dE, \qquad (4a)$$

where $f(x) = \frac{1}{1+\exp(x/k_B T)}$ is the Fermi-Dirac distribution function, $k_B = 1.3807 \times 10^{-23}$ J/K, $T$ is the absolute temperature, and $g_G^\pm(E)$ is the 2D-density of states (DOS). The DOS, derived in **Appendix A.2**, is

$$g_G^\pm(E) = \sum_s \int_{-\infty}^{+\infty} \frac{q \, dq}{\pi} \delta[E - E_s^\pm(q)] = \sum_s \frac{|E - D_0 - D_s|}{\pi \hbar^2 v_s^2} H\left(\pm(E - D_0 - D_s) - \frac{\Delta_s}{2}\right), \qquad (4b)$$

where $q = \sqrt{q_x^2 + q_y^2}$; and $H(E)$ is the Heaviside step-function, $H(E > 0) = 1$ and $H(E < 0) = 0$.

Substitution of DOS (4b) into the Eqs.(4a) and integration lead to the expression (see **Appendix A.2**):

$$n(E_F) = \sum_s \left( \frac{\Delta_s k_B T}{2\pi \hbar^2 v_s^2} \ln\left[1 + e^{\frac{E_F - D_0 - D_s - \frac{\Delta_s}{2}}{k_B T}}\right] - \frac{(k_B T)^2}{\pi \hbar^2 v_s^2} \mathrm{Li}_2\left[-e^{\frac{E_F - D_0 - D_s - \frac{\Delta_s}{2}}{k_B T}}\right]\right), \qquad (4c)$$

where $\mathrm{Li}_2[x]$ is a particular case of the polylogarithm function $\mathrm{Li}_m[x] = \sum_{k=1}^\infty \frac{x^k}{k^m}$. Expression for the holes can be obtained from Eq.(4c) with the substitution $E_F \to -E_F, D_0 \to -D_0$ and $D_s \to$ $-D_s$. Note that an exact analytical expression (4c) shows how the two-dimensional concentration of carriers depends on the proximity effect (via the combination of parameters $D_0 - D_s - \frac{\Delta_s}{2}$), and on the Fermi level $E_F$. To the best of our knowledge, any expression of (4c) type was not derived earlier.

For the temperatures $k_B T \ll \left| E_F - D_0 - D_s \mp \frac{\Delta_s}{2} \right|$ we the graphene charge density can be estimated as:

$$e[p(E_F) - n(E_F)] \approx \frac{e}{2\pi h^2} \sum_s \frac{1}{v_s^2} \left( (E_F - D_0 - D_s)^2 - \left(\frac{\Delta_s}{2}\right)^2 \right) \text{sign}(D_0 + D_s - E_F). \quad (5)$$

It appeared that Eq.(5) is almost accurate for $Y_3Fe_5O_{12}/Gr$ and $CoFeO_4/Gr$ parameters at temperatures $T<(350 - 400)$ K, otherwise we will solve Eq.(4c) for the Fermi level determination. Notably that Eq.(5) allows us to express the Fermi level $E_F$ via the graphene charge density in a self-consistent manner.

Below we assume that $M_s(E_F) = 0$ when a Fermi level is inside the energy gap of the spectrum (2). Outside the gap $M_s(E_F)$ is described by the expression [18]:

$$M_s = \text{int}\left[\frac{2W}{\lambda_{DB}^s}\right], \quad (6a)$$

where the symbol "int" denotes the integer part, and $\lambda_{DB}^s$ is the electron de Broglie wavelength:

$$\lambda_{DB}^s = \frac{2\pi}{q_s(E_F)} \approx \frac{2\pi \hbar v_s}{\sqrt{(E_F - D_0 - D_s)^2 - (\Delta_s/2)^2}}, \quad (6b)$$

where $q_s(E) = \frac{1}{\hbar v_s}\sqrt{(E - D_0 - D_s)^2 - (\Delta_s/2)^2}$ is solution of Eq.(2b) for the given energy $E$; and the approximate equality is valid under the assumption $v_s \approx v_F$.

The physical meaning of $M(E_F)$ for 2D channel is the number of de Broglie half-wavelengths which can be located at the width of this channel $W$.

Finally let's analyze the transmission coefficient $T_s(E_F)$ introduced in Ref.[10]. We use here a standard Mott's two-channel model for conductivity [19] and moreover assume the hierarchy of lengths $\lambda_{AP} \ll L \ll \lambda_P$, where $\lambda_{AP}, \lambda_P$ is electron mean free path for spin polarization anti-parallel (parallel) to spin polarization of spin majority for a proper position of Fermi energy. This yields in the first approximation order.

For spins "**up**" $T_\uparrow(E_F) \approx 1$ if the Fermi level $E_F$ satisfies the inequalities $E_F < D_0 + D_\uparrow - \frac{\Delta_\uparrow}{2}$ or $E_F > D_0 + D_\uparrow + \frac{\Delta_\uparrow}{2}$; and $T_\uparrow(E_F) \approx 0$ if the Fermi level satisfies the inequality $D_0 + D_\uparrow - \frac{\Delta_\uparrow}{2} < E_F < D_0 + D_\uparrow + \frac{\Delta_\uparrow}{2}$.

For spin "**down**" $T_↓(E_F) ≈ 1$ if the Fermi level $E_f$ satisfies the inequalities $E_F < D_0 + D_↓ - \frac{Δ_↓}{2}$ or $E_F > D_0 + D_↓ + \frac{Δ_↓}{2}$; and $T_↓(E_F) ≈ 0$ if the Fermi level satisfies the inequality $D_0 + D_↓ - \frac{Δ_↓}{2} < E_F < D_0 + D_↓ + \frac{Δ_↓}{2}$.

### E. Influence of the ferroelectric polarization on the 2D carriers in graphene

Next step is to define the density $n$ of 2D carriers in graphene in a self-consistent way. Since we consider a single-domain ferroelectric film, where the electric field is constant in, the surface charge $σ$ stored in graphene is [20]:

$$σ = \frac{h\, P_3}{h + d_{eff}}, \quad (7a)$$

where we introduced an effective width of dielectric gap,

$$d_{eff} = ε_f \left[\frac{d}{ε_d} + \frac{l_s}{ε_G} \tanh\left(\frac{g}{l_s}\right)\right], \quad (7b)$$

and regard that a single-layer graphene of effective thickness $g \sim 0.312$ nm can be characterized by the effective screening length $l_S$ and dielectric permittivity $ε_G \sim 1$ (in the normal direction). The background permittivity [21] of a ferroelectric film in the out-of-plane direction is $ε_f \sim 10$. Since the length $l_S$ of a semi-metal is usually smaller (or significantly smaller) than 0.1 nm [21, 22], we obtain that the strong inequality $l_S \ll 2g$ is characteristic for a gapless semiconductor. A ferroelectric film has a thickness $h \gg 2g \gg d$ and so $h \gg l_S$. Thus, the concentration of free carriers in graphene acquires the form [20]:

$$p - n = \frac{σ}{e} = \frac{h}{h + d_{eff}} \frac{P_3}{e}. \quad (8)$$

where $e = 1.6 × 10^{-19}$ C is an electron charge and $d_{eff} ≈ ε_f \left(\frac{d}{ε_d} + \frac{l_s}{ε_G}\right)$. For numerical estimates of $d_{eff}$ we can regard that an ultra-thin dielectric gap has a thickness $d<0.1$ nm and dielectric permittivity $ε_d \cong 1$.

### III. INFLUENCE OF MISFIT STRAIN AND TEMPERATURE ON THE FERMI ENERGY

Within continuous media Landau-Ginzburg-Devonshire (LGD) approach [23], the value and orientation of the spontaneous polarization $P_i$ in thin ferroelectric films can be controlled by size effect, temperature $T$ and misfit strain $u_m$ originated from the film-substrate lattice constants mismatch [24, 25]. The density of LGD free energy, which minimization allows to calculate phase diagram and polarization, has the form:

$$g_{LGD} = a_1(P_1^2 + P_2^2) + a_3 P_3^2 + a_{11}(P_1^4 + P_2^4) + a_{33} P_3^4 + a_{12} P_1^2 P_2^2 + a_{13}(P_1^2 + P_2^2)P_3^2 +$$
$$a_{111}(P_1^6 + P_2^6 + P_3^6) + a_{112}[P_1^2(P_2^4 + P_3^4) + P_2^2(P_1^4 + P_3^4) + P_3^2(P_2^4 + P_1^4)] + a_{123} P_1^2 P_2^2 P_3^2 \quad (9a)$$

The coefficients:

$$a_1 = \alpha_{1T}(T - T_C^f) - \frac{(Q_{11}+Q_{12})u_m}{s_{11}+s_{12}}, \quad a_3 = \alpha_{1T}(T - T_C^f) - \frac{2Q_{12}u_m}{s_{11}+s_{12}} + \frac{d_{eff}}{\varepsilon_0\varepsilon_f(h+d_{eff})}, \tag{9b}$$

$$a_{11} = \alpha_{11} + \frac{s_{11}(Q_{11}^2+Q_{12}^2)-2Q_{11}Q_{12}s_{12}}{2(s_{11}^2-s_{12}^2)}, \quad a_{33} = \alpha_{11} + \frac{Q_{12}^2}{s_{11}+s_{12}}, \tag{9c}$$

$$a_{12} = \alpha_{12} - \frac{s_{12}(Q_{11}^2+Q_{12}^2)-2Q_{11}Q_{12}s_{11}}{s_{11}^2-s_{12}^2}, \quad a_{13} = \alpha_{12} + \frac{Q_{12}(Q_{11}+Q_{12})}{s_{11}+s_{12}}. \tag{9d}$$

Here $T_C^f$ is the Curie temperature of bulk ferroelectric, $Q_{ij}$ are the components of electrostriction tensor, $s_{ij}$ are elastic compliances. The positive coefficient $\alpha_T$ is proportional to the inverse Curie-Weiss constant. When deriving Eq.(9b) we used the expression for the depolarization field inside the ferroelectric film, $E_3 = -\frac{P_3}{\varepsilon_0\varepsilon_f}\frac{d_{eff}}{h+d_{eff}}$.

The phase diagram of the 200-nm PbZr$_{0.4}$Ti$_{0.6}$O$_3$ (**PZT**) film in coordinates temperature – misfit strain, which contains the paraelectric (PE) and ferroelectric (FE) phases the with out-of-plane (FE$_C$), in-plane (FE$a$) and mixed (FE$r$) orientation of polarization vector, is shown in **Fig. 3a.** The dependence of the out-of-plane polarization on the misfit strain and temperature is shown in **Fig. 3b**.

Note that the phase diagram in **Fig. 3a** differs from the diagrams calculated by Pertsev et al. [14] by the presence of finite size effect. The size effect scale is determined by the ratio $d_{eff}/h$ [see Eq.(9b)] and we put $h$=200 nm and $d_{eff}$=1 nm in **Fig. 3.**

The boundaries between FE$_C$, FE$r$ and FE$a$ are calculated numerically. The boundaries of PE phase instability are described by the analytical expressions $a_1 = 0$ and $a_3 = 0$, which allow to estimate PE-FEc and PE-FEa transition temperatures as:

$$T_{PE-FE_c}(u_m, h) \approx T_C^f\left(1 + \frac{2Q_{12}u_m}{\alpha_T T_C(s_{11}+s_{12})}\right) - \frac{1}{\alpha_T\varepsilon_0\varepsilon_f}\frac{d_{eff}}{h+d_{eff}}, \tag{10a}$$

$$T_{PE-FE_a}(u_m, h) \approx T_C^f\left(1 + \frac{(Q_{11}+Q_{12})u_m}{\alpha_T T_C(s_{11}+s_{12})}\right). \tag{10b}$$

Note that only out-of-plane component of the spontaneous polarization $P_3$ can induce the significant number of carriers in graphene [see Eq.(8)], and the in-plane components $P_1$ and $P_2$ do not influence the carrier density. So, one we should select those misfit strains $u_m$ and film thicknesses $h$, for which the FE$_c$ or FE$_r$ phases with nonzero $P_3$ are stable at the working temperature. Meanwhile PE and FE$_a$ phases with $P_3 = 0$, do not present any interest for the spin-polarization effect.

It is seen from **Fig. 3a-3b** that PZT film should be strained with $u_m < 0.1\%$ in order to increase the region of stable FE$_c$ or FE$_r$ phases up to room and elevated temperatures. Also, the temperature stability of FE$c$ and FE$r$ phases increases with the strain decrease, and becomes highest at $u_m < -1\%$. This happens because the compressive strain ($u_m < 0$) increases $T_{PE-FE_c}(u_m, h)$, and tensile strain ($u_m > 0$) decreases it, since $Q_{12} < 0$ and $s_{11} + s_{12} > 0$ for PZT (see **Table II**).

Only the out-of-plane component of polarization, $P_3(T, u_m, h)$, can change the Fermi energy, and the dependence of $E_F$ on $P_3$ is given by expression

Under the temperature limiting conditions and assuming that $v_s \approx v_F$, the approximate expression for Fermi energy should be found from the equation $e[p(E_F) - n(E_F)] + \sigma = 0$:

$$E_F(\sigma) \approx D_0 + \frac{D_\uparrow + D_\downarrow}{2} + \sqrt{\frac{\Delta_\uparrow^2 + \Delta_\downarrow^2}{8} - \left(\frac{D_\uparrow - D_\downarrow}{2}\right)^2 + \frac{\pi \hbar^2 v_F^2 \sigma}{e}} \qquad (11)$$

Which is valid when $|E_F - D_0 - D_\uparrow| \gg \frac{\Delta_\uparrow}{2}$ and $|E_F - D_0 - D_\downarrow| \gg \frac{\Delta_\downarrow}{2}$, and also $\sigma = \frac{h\, P_3(T, u_m, h)}{h + d_{eff}}$ in accordance with Eq.(7a).

Note that the expression (11) significantly benefits from the expression $E_F = \hbar v_F \sqrt{\pi n}$, used in Ref.[13] for the case, since it accounts for the proximity effect, polarization rotation and depolarization effect in the presence of dielectric gap (7b). Not less important is that, since the charge density $\sigma$ is defined by the strain-dependent polarization $P_3(T, u_m, h)$, analytical expression (11) shows that the ferroelectric polarization governs the Fermi level $E_F(\sigma)$ in a self-consistent manner. To the best of our knowledge Eq.(11) was not derived earlier.

The dependences of the Fermi energy and de Broglie wavelength on the misfit strain and temperature are shown in **Fig. 3c** and **3d**, respectively. It is seen that misfit strain makes it possible to control $E_F$ and $\lambda_{DB}$ values in FE$_C$ or FE$r$ phases up to room and higher temperatures; at that the temperature range of the phases increases with compressive strain increase. Actually, we can vary the values of $E_F$ and $\lambda_{DB}$ in the range $(0 - 2.3)$eV and $(0 - 5)$nm, respectively, by a proper choice of $u_m$ and $T$ (see the gradient color regions of FE$_C$ and FE$r$ phases, which are separated by a slightly visible boundary in **Fig. 3c** and **3d**). The triangle-like violet region in **Fig. 3c** and the white region in **Fig. 3d** are the PE or FE$a$ phases, where the in-plane ferroelectric polarization corresponding to FE$a$ phase (or its absence in PE phase) cannot change $E_F$ and $\lambda_{DB}$ in a graphene. It should be noted that the difference between $\lambda_{DB}^\downarrow$ and $\lambda_{DB}^\uparrow$ turned out to be insignificant for both considered nanostructures (less than 10%), therefore, **Fig. 3d** shows only one of these two quantities.

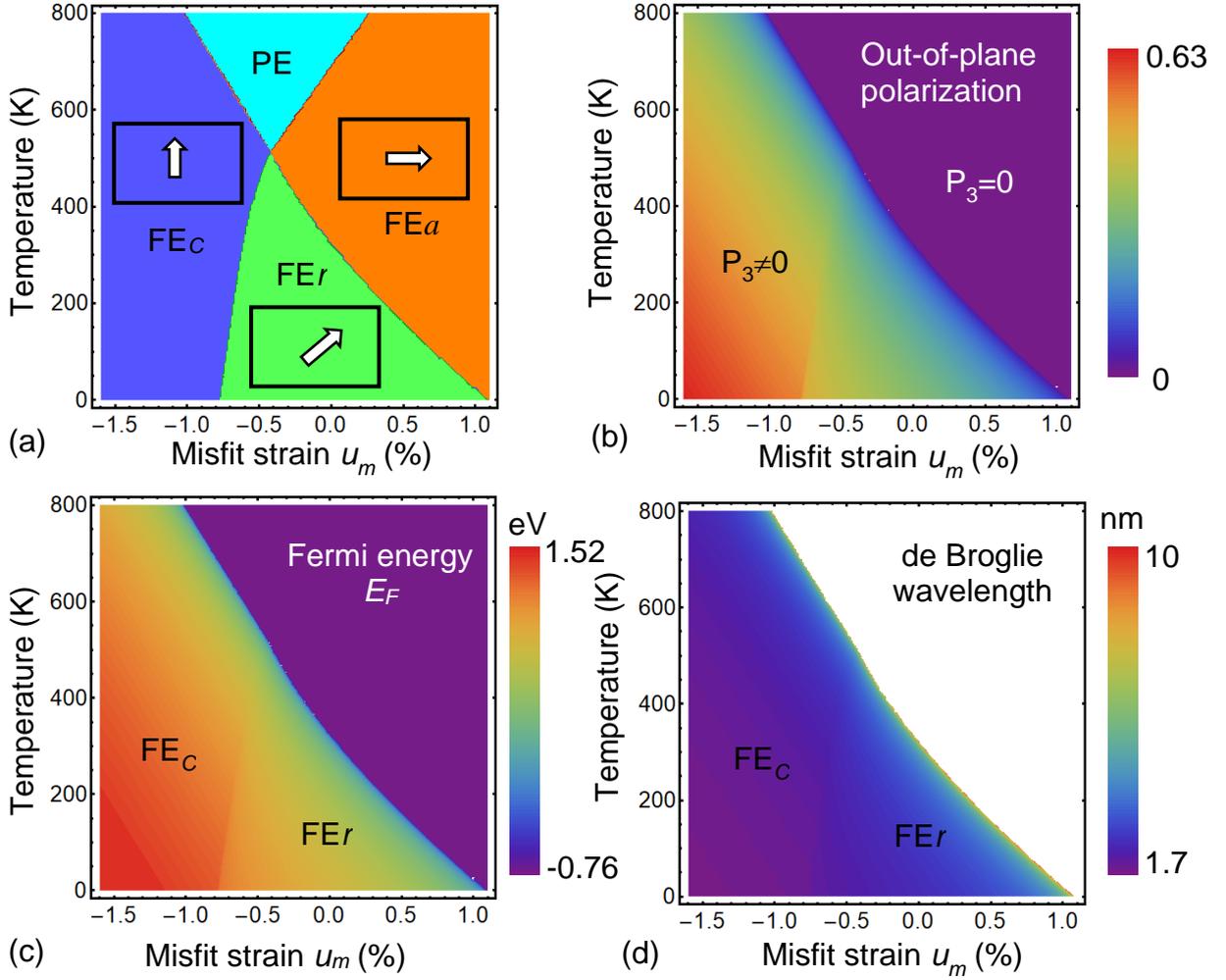

**FIGURE 3**. (a) The phase diagram of the PZT film polarization in coordinates temperature – misfit strain, which contains the paraelectric (PE) and ferroelectric phases with the out-of-plane (FE$_c$), in-plane (FE$_a$) and mixed (FE$_r$) orientation of polarization vector. (**b-d**) The dependences of the out-of-plane polarization (**b**), Fermi energy (**c**) and de Broglie wavelength (**d**) on the misfit strain and temperature. Color bars show the range of corresponding values at the contour maps; white arrows inside empty rectangles show the polarization direction in the film. The film thickness $h$=200 nm and effective gap width $d_{eff}$=1 nm. LGD parameters are listed in **Table II**.

**Table II.** Parameters of PbZr$_{0.4}$Ti$_{0.6}$O$_3$, used in LGD phenomenological modelling from [26]

| $\alpha_{1T}$ | $T_C^f$ | $\alpha_{11}$ | $\alpha_{12}$ | $a_{111}$ | $a_{112}$ | $a_{123}$ | $Q_{11}$ | $Q_{12}$ | $Q_{44}$ |
|---|---|---|---|---|---|---|---|---|---|
| $10^5 \frac{m}{FK}$ | °C | $10^7 \frac{m^5}{FC^2}$ | $10^8 \frac{m^5}{FC^2}$ | $10^8 \frac{m^9}{FC^4}$ | $10^8 \frac{m^9}{FC^4}$ | $10^9 \frac{m^9}{FC^4}$ | $10^2 \frac{m^4}{C^2}$ | $10^2 \frac{m^4}{C^2}$ | $10^2 \frac{m^4}{C^2}$ |
| 2.121 | 418.4 | 3.614 | 3.233 | 1.859 | 8.503 | −4.063 | 8.116 | -2.950 | 6.710 |

# IV. THE SPIN-DEPENDENT CONDUCTANCE OF FERROMAGNETIC-GRAPHENE-FERROELECTRIC NANOSTRUCTURE

The spin-polarization coefficient $p$ is defined as [10]

$$p = \frac{G_\uparrow - G_\downarrow}{G_\uparrow + G_\downarrow} \qquad (12)$$

The temperature dependence of the spin "up" and "down" graphene channel conductivities, $G_\downarrow$ and $G_\uparrow$, and spin polarization $p$ were calculated for CoFeO$_4$/graphene/PZT [**Fig. 4a-c**] and Y$_3$Fe$_5$O$_{12}$/graphene/PZT [**Fig. 4d-f**] nanostructures. The curves 1 - 3 in **Figs. 4** correspond to three different values of misfit strain. The temperature dependence of the conductivities $G_\downarrow$ and $G_\uparrow$ are step-like in the temperature range corresponding to the FE phase of the 200-nm PZT film. The spin-polarization effect disappears in FE*a* and PE phases. The steps become longer with temperature decrease.

One can see that CoFeO$_4$/graphene/PZT system demonstrates very narrow temperature windows for the value $p = \pm 1$ (100% spin polarization). In fact, these values are pinned to the temperature ranges of fraction of Kelvin order width. This can make operation of the system as a spin filter unstable; however, by controlling the temperature shift within several K range we can get a promising spin valve, which switches at the defined temperature from 0% to 100% spin polarization, and almost immediately to 0% polarization again. Zero values of the spin polarization $p$ are shown in the log scale in **Fig. 4c**, since the temperature dependence of $p$ has the form of a step function, and $p$ is equal to zero in some temperature ranges. On the contrary, the Y$_3$Fe$_5$O$_{12}$/graphene/PZT system demonstrates the absence of the temperature windows with value $p = 0$, which makes it promising both as a stable spin filter and as a spin valve. It should be noted that the sign of $p$ is defined by the sign of the spontaneous polarization in the single-domain ferroelectric film. The fact can open unexplored possibilities to modulate *p* using a ferroelectric film with a domain structure, which is not considered in this work.

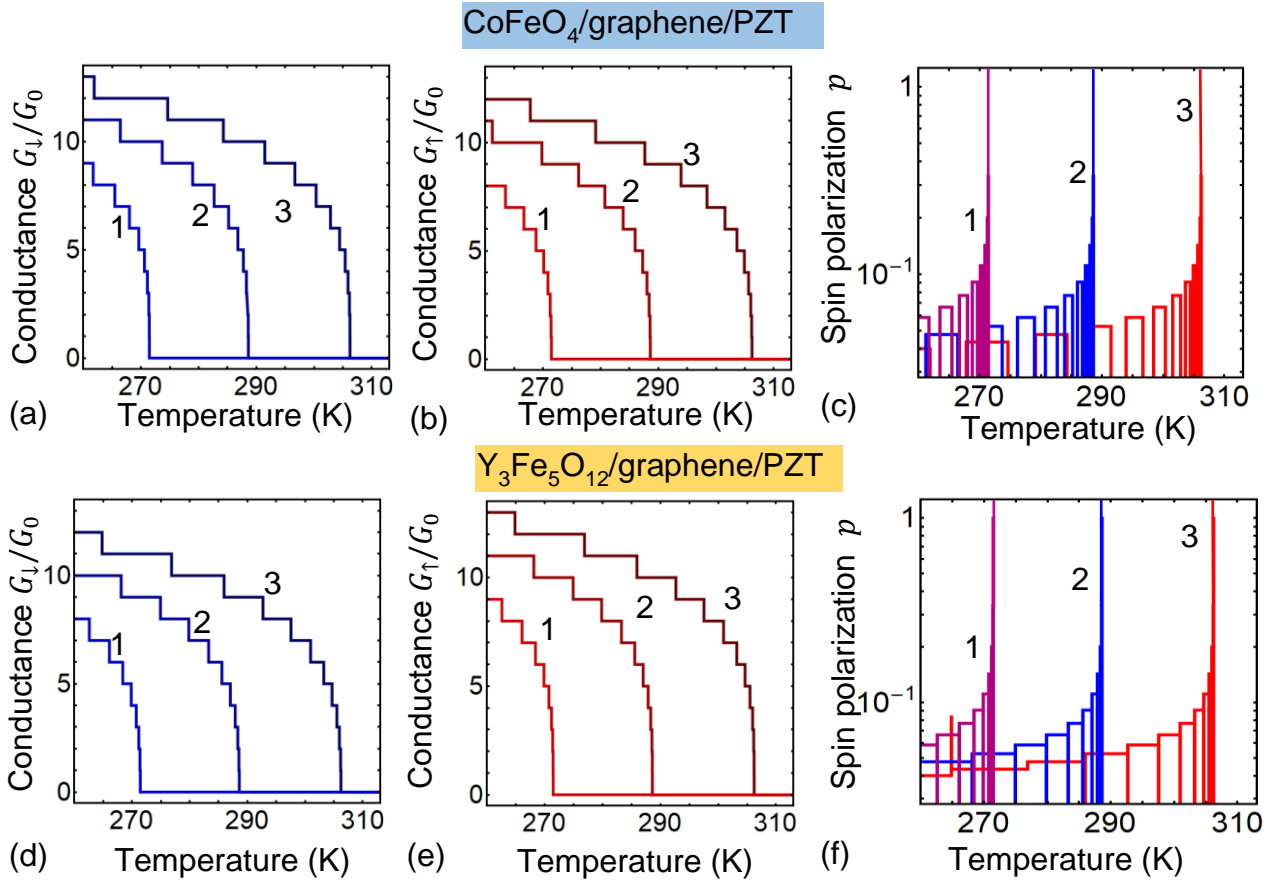

**FIGURE 4.** The temperature dependence of the graphene channel conductivities $G_\downarrow$ **(a, d)** and $G_\uparrow$ **(b, e),** and absolute value of spin polarization $p$ **(c, f)** calculated for CoFeO$_4$/graphene/PZT **(a, b, c)** and Y$_3$Fe$_5$O$_{12}$/graphene/PZT **(d, e, f)** nanostructures. The curves 1-3 correspond to several values of misfit strain, namely $u_m$=1.5, 1, 0.5 ×10$^{-3}$. The width of the graphene channel $W$ = 50 nm; the film thickness is $h$=200 nm, and the gap width $d_{eff}$=1 nm. The conductance is normalized on $G_0 = e^2/2\pi\hbar$. Parameters for the energy levels are listed in **Table I**. Ferroelectric parameters are listed in **Table II**.

### B. Strain engineering of the spin-polarization

The dependence of the spin-polarization $p$ on temperature and misfit strain calculated for CoFeO$_4$/graphene/PZT and Y$_3$Fe$_5$O$_{12}$/graphene/PZT nanostructures are shown in **Fig. 5a** and **5b**, respectively. For both nanostructures one can see two triangle-like regions with $p = 0$ separated by a thin line-like region with maximal spin-polarization $|p| = 1$. To reach $p = \pm 1$ at room temperature the strain $u_m \approx (0.07 - 0.09)\%$ should be applied to a 200-nm thick PZT film. The strain becomes smaller at higher temperatures. The region with spin polarization $p \to 1$ is very narrow in the phase diagram, and it is localized near the vanishing curve of the out-of-plane component of ferroelectric polarization $P_3$. Far from this curve the values of spin polarization are of the order of 0.1.

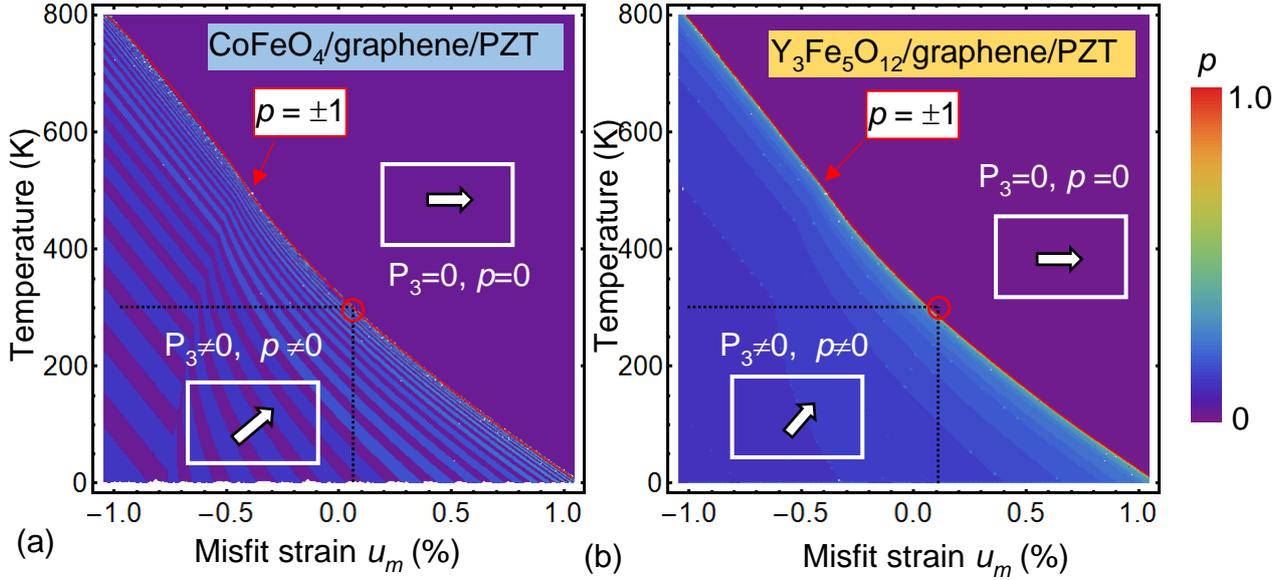

**FIGURE 5.** Spin polarization $p$ in coordinates temperature – misfit strain calculated for **(a)** CoFeO$_4$/graphene/PZT and **(b)** Y$_3$Fe$_5$O$_{12}$/graphene/PZT nanostructures. Parameters are the same as in **Fig. 3-4**.

In fact, **Fig. 5** demonstrates the realistic possibilities to control the spin-polarized conductance of graphene by a misfit strain ("strain engineering") from low (< 100 K) up to room (~ 300 K) and higher temperatures (350 – 550) K in the nanostructures CoFeO$_4$/graphene/PZT and Y$_3$Fe$_5$O$_{12}$/graphene/PZT. This makes systems under consideration promising candidates for the fabrication of novel type spin valves and spin filters.

## V. CONCLUSION

We calculated a spin-polarized conductance in the nanostructure "high temperature ferromagnetic insulator/ graphene/ ferroelectric film" with a special attention to the impact of electric polarization rotation in a strained multiaxial ferroelectric film. The rotation and value of polarization vector are controlled by a misfit strain.

The proposed phenomenological model takes into account the shift of the Dirac point induced by the proximity of ferromagnetic insulator. We use the Landauer formula for the conductivity of the graphene channel, where the strain-dependent ferroelectric polarization governs the concentration of two-dimensional charge carriers and Fermi level in graphene in a self-consistent manner.

Mention the essential restriction imposed on the graphene channel length in such device: it should be shorter than an electron mean free path for spin polarization parallel with the one in the graphene channel modified by ferromagnetic insulator, and longer than an electron mean free path for spin polarization antiparallel with the one in graphene channel. However, because of long spin-flip length in a standard graphene-on-substrate (up to several μm scales), this restriction doesn't lead to ultra-short channels and seems to be not very critical.

We demonstrated the real opportunities to control the spin-polarized conductance of graphene by a misfit strain ("strain engineering") at room and higher than room temperatures in the nanostructures $CoFeO_4$/graphene/PZT and $Y_3Fe_5O_{12}$/graphene/PZT. Namely, temperature and strain ranges exist, where spin polarization of the systems under consideration can switch from 0% to 100%, and then to 0% again. Obtained results open the possibilities for the applications of ferromagnetic/graphene/ferroelectric nanostructures as non-volatile spin filters and spin valves operating at room and higher temperatures.

**Authors' contribution.** E.A.E. wrote the codes for results visualization and made figures. A.N.M. generated the research idea, performed analytical calculations and wrote the manuscript draft. M.V.S. contributed to the analytical calculations, results discussion and manuscript improvement.

A.N.M. expresses her deepest gratitude to Prof. S.M. Ryabchenko (NASU) for critical remarks and stimulating discussions. This project has received funding from the European Union's Horizon 2020 research and innovation programme under the Marie Skłodowska-Curie grant agreement No 778070.

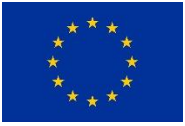

## APPENDIX A

**A.1. Parameters used in Eq.(2c).** Parameters used in Eq.(2c) and listed in **Table I** for several FM-Gr pairs have been recalculated from Hallal et al. by the following procedure. Hallal et al. used the effective Hamiltonian [9]

$$\widehat{H}_s^{\pm}(q) = \hbar v_s \widehat{\boldsymbol{\sigma}} \boldsymbol{q} \boldsymbol{I}_s + \frac{\delta}{2} \boldsymbol{I}_s \hat{s}_z + \frac{\Delta_s}{2} \hat{\sigma}_z \boldsymbol{I}_s + \frac{\Delta_\delta}{2} \hat{\sigma}_z \hat{s}_z, \qquad (A.1)$$

where $\boldsymbol{\sigma}$ and $s$ are the Pauli matrices that act on the sublattice and spin, respectively. The second term represents the exchange coupling induced by the magnetic moment of magnetic atoms, with $\delta = \frac{\delta_e + \delta_h}{2}$, where $\delta_h$ and $\delta_e$ are the strength of exchange spin-splitting of the hole and electron, respectively. The third term results from the fact that graphene sublattices *A* and *B* are now feeling different potential, which result in a spin-dependent band gap opening at the Dirac point $\Delta_s = \frac{\Delta_\uparrow + \Delta_\downarrow}{2}$. The fourth term is proportional to the spin-dependent "mass-constant" $\Delta_\delta = \frac{\delta_e - \delta_h}{2} = \frac{\Delta_\uparrow - \Delta_\downarrow}{2}$.

The characteristic energies of the band edges $(E_s^\pm(q) - D_0)$ arising in the band spectrum at $q=0$ in the proximity of a ferromagnetic insulator are $D_\uparrow - \frac{\Delta_\uparrow}{2}, D_\uparrow + \frac{\Delta_\uparrow}{2}$, for spins "up" and: $D_\downarrow - \frac{\Delta_\downarrow}{2}$, $D_\downarrow + \frac{\Delta_\downarrow}{2}$ for spins "down". Thus $\left(D_\uparrow + \frac{\Delta_\uparrow}{2}\right) - \left(D_\downarrow + \frac{\Delta_\downarrow}{2}\right) = \delta_e$, and $\left(D_\uparrow - \frac{\Delta_\uparrow}{2}\right) - \left(D_\downarrow - \frac{\Delta_\downarrow}{2}\right) = \delta_h$. Hence $2(D_\uparrow - D_\downarrow) = 4D_\uparrow = \delta_e + \delta_h$, $\Delta_\uparrow - \Delta_\downarrow = \delta_e - \delta_h$ and so $D_{\uparrow,\downarrow} = \pm\frac{\delta_e + \delta_h}{4}$.

Effective Hamiltonian of Song et al. is:

$$\widehat{H}_s^\pm(q) = \hat{\sigma}_0(D_0 + D_s) + (\hbar v_s \hat{\boldsymbol{\sigma}} \boldsymbol{q} \pm \Delta_s \hat{\sigma}_z), \tag{A.2a}$$

where $\hat{\sigma}_0 = \begin{pmatrix} 1 & 0 \\ 0 & 1 \end{pmatrix}$, $\hat{\boldsymbol{\sigma}} = (\hat{\sigma}_x, \hat{\sigma}_y)$, $\hat{\sigma}_x = \begin{pmatrix} 0 & 1 \\ 1 & 0 \end{pmatrix}$, $\hat{\sigma}_y = \begin{pmatrix} 0 & -i \\ i & 0 \end{pmatrix}$ and $\hat{\sigma}_z = \begin{pmatrix} 1 & 0 \\ 0 & -1 \end{pmatrix}$. Since $\boldsymbol{q} = (q_x, q_y)$, we obtained that the scalar product $\hat{\boldsymbol{\sigma}} \boldsymbol{q} = \begin{pmatrix} 0 & q_x - iq_y \\ q_x + iq_y & 0 \end{pmatrix}$ and $\Delta_s \hat{\sigma}_z = \begin{pmatrix} \Delta_s & 0 \\ 0 & -\Delta_s \end{pmatrix}$.

The analytical dependence for the energy levels is:

$$E_s^\pm(q) = D_0 + D_s \pm \sqrt{(\hbar v_s q)^2 + (\Delta_s/2)^2}. \tag{A.2b}$$

Corresponding effective "masses" of graphene carriers, $m_s$, can be found from the expansion of expression (A.2b) at small $\boldsymbol{q}$, $E_s^\pm(q) - D_0 - D_s \approx \pm \frac{\Delta_s}{2}\left(1 + \frac{1}{2}\left(\frac{2\hbar v_s q}{\Delta_s}\right)^2\right) = \pm\frac{\Delta_s}{2} \pm \frac{2v_s^2}{\Delta_s}\frac{(\hbar q)^2}{2}$, as $m_s = \mp\frac{\Delta_s}{2v_s^2}$.

**A.2. DOS derivation**

$$g_G^\pm(E) = 2\sum_s \int_{-\infty}^{+\infty} \frac{dq_x dq_y}{(2\pi)^2}\delta[E - E_s^\pm(q)] \equiv \sum_s \int_0^{+\infty} \frac{q dq}{\pi}\delta[E - E_s^\pm(q)], \tag{A.3a}$$

where $\boldsymbol{q} = \{q_x, q_y\}$ and $q = \sqrt{q_x^2 + q_y^2}$, and degeneracy of K and K' values gives the factor "2". Using the property $\int_{-\infty}^{+\infty}\delta[a(q)]b(q)dq = \sum_i \frac{b(q_i)}{|a'(q_i)|}$ of the Dirac-delta function $\delta[q]$, where $a'(q) = \frac{da}{dq}$, is the derivative, we obtained from Eq.(4b) that

$$g_G^+(E) = \sum_s \frac{q_s}{\pi}\left|\frac{dq_s}{dE_s^+}\right| H\left(E - D_0 - D_s - \frac{\Delta_s}{2}\right), \tag{A.3b}$$

$$g_G^-(E) = \sum_s \frac{q_s}{\pi}\left|\frac{dq_s}{dE_s^-}\right| H\left(-E + D_0 + D_s - \frac{\Delta_s}{2}\right). \tag{A.3c}$$

Where $q_s(E) = \frac{1}{\hbar v_s}\sqrt{(E - D_0 - D_s)^2 - (\Delta_s/2)^2}$ is the positive solution of Eq.(A.2b) for the given energy $E$; and $H(E)$ is the Heaviside step-function, $H(E > 0) = 1$ and $H(E < 0) = 0$. The step-function appears in DOS expression since the equations $E = E_s^+(q)$ and $E = E_s^-(q)$ have the physical solution $q > 0$ only for the cases $E > D_0 + D_s + \frac{\Delta_s}{2}$ and $E < D_0 + D_s - \frac{\Delta_s}{2}$, respectively.

Using the expressions for the derivatives $\frac{dE_s^\pm(q)}{dq} = \frac{\pm(\hbar v_s)^2 q}{\sqrt{(\hbar v_s q)^2 + (\Delta_s/2)^2}}\bigg|_{q=q_s} \equiv \frac{(\hbar v_s)^2 q_s}{E - D_0 - D_s}$, or the inverse derivative $\frac{dq_s}{dE} = \frac{E - D_0 - D_s}{(\hbar v_s)^2 q_s}$, we obtained from Eq.(A.3b) the following expressions:

$$g_G^+(E) = \sum_s \frac{|E-D_0-D_s|}{\pi \hbar^2 v_s^2} H\left(E - D_0 - D_s - \frac{\Delta_s}{2}\right) \equiv \sum_s \frac{\sqrt{(\hbar v_s q_s)^2 + (\Delta_s/2)^2}}{\pi \hbar^2 v_s^2} H\left(E - D_0 - D_s - \frac{\Delta_s}{2}\right), \quad \text{(A.4a)}$$

$$g_G^-(E) = \sum_s \frac{|E-D_0-D_s|}{\pi \hbar^2 v_s^2} H\left(D_0 + D_s - \frac{\Delta_s}{2} - E\right) \equiv \sum_s \frac{\sqrt{(\hbar v_s q_s)^2 + (\Delta_s/2)^2}}{\pi \hbar^2 v_s^2} H\left(D_0 + D_s - \frac{\Delta_s}{2} - E\right). \quad \text{(A.4b)}$$

Now the 2D-concentrations of electrons $n$ and holes $p$ could be calculated as follows:

$$n(E_F) = \int_{-\infty}^{+\infty} g_G^+(E) f(E - E_F) dE = \sum_s \int_{-\infty}^{+\infty} \frac{|E-D_0-D_s|}{\pi \hbar^2 v_s^2} H\left(E - D_0 - D_s - \frac{\Delta_s}{2}\right) f(E - E_F) dE = \sum_s \int_{D_0+D_s+\frac{\Delta_s}{2}}^{+\infty} \frac{|E-D_0-D_s|}{\pi \hbar^2 v_s^2 \left(1+\exp\left(\frac{E-E_F}{k_B T}\right)\right)} dE = \left|E - D_0 - D_s - \frac{\Delta_s}{2} \equiv \tilde{E}\right| =$$

$$\sum_s \int_0^{+\infty} \frac{\tilde{E} + \frac{\Delta_s}{2}}{\pi \hbar^2 v_s^2 \left(1+\exp\left(\frac{\tilde{E}-E_F+D_0+D_s+\frac{\Delta_s}{2}}{k_B T}\right)\right)} d\tilde{E} = \sum_s \left[\frac{\Delta_s k_B T}{2\pi \hbar^2 v_s^2} \ln\left(1 + e^{\frac{E_F-D_0-D_s-\frac{\Delta_s}{2}}{k_B T}}\right) - \frac{(k_B T)^2}{\pi \hbar^2 v_s^2} \text{Li}_2\left(-e^{\frac{E_F-D_0-D_s-\frac{\Delta_s}{2}}{k_B T}}\right)\right] \quad \text{(A.5a)}$$

$$p(E_F) = \int_{-\infty}^{+\infty} g_G^-(E) f(E_F - E) dE = \int_{-\infty}^{+\infty} \sum_s \frac{|E-D_0-D_s|}{\pi \hbar^2 v_s^2} H\left(D_0 + D_s - \frac{\Delta_s}{2} - E\right) f(E_F - E) dE =$$

$$\sum_s \int_{-\infty}^{D_0+D_s-\frac{\Delta_s}{2}} \frac{|E-D_0-D_s|}{\pi \hbar^2 v_s^2 \left(1+\exp\left(\frac{E_F-E}{k_B T}\right)\right)} dE = \left|D_0 + D_s - \frac{\Delta_s}{2} - E \equiv \tilde{E}\right| ==$$

$$\sum_s \int_0^{+\infty} \frac{\tilde{E} + \frac{\Delta_s}{2}}{\pi \hbar^2 v_s^2 \left(1+\exp\left(\frac{\tilde{E}+E_F-D_0-D_s+\frac{\Delta_s}{2}}{k_B T}\right)\right)} d\tilde{E} = \sum_s \left[\frac{\Delta_s k_B T}{2\pi \hbar^2 v_s^2} \ln\left(1 + e^{\frac{-E_F+D_0+D_s-\frac{\Delta_s}{2}}{k_B T}}\right) - \frac{(k_B T)^2}{\pi \hbar^2 v_s^2} \text{Li}_2\left(-e^{\frac{-E_F+D_0+D_s-\frac{\Delta_s}{2}}{k_B T}}\right)\right] \quad \text{(A.5b)}$$

Here we suppose that $\Delta_s > 0$, $\ln(x) \equiv \log_e(x)$ and $\text{Li}_2[x]$ is a particular case of the polylogarithm function $\text{Li}_m[x] = \sum_{k=1}^{\infty} \frac{x^k}{k^m}$. These very cumbersome expressions could be essentially simplified in two limiting cases. Namely, we recall the approximate relations, valid in different asymptotic cases:

$$-\text{Li}_2(-e^\xi) \cong \begin{cases} e^\xi & \text{at } \xi < 0 \text{ and } |\xi| \gg 1, \\ \frac{\xi^2}{2} & \text{at } \xi > 0 \text{ and } |\xi| \gg 1. \end{cases} \quad \text{(A.6a)}$$

$$\ln(1 + e^\xi) \cong \begin{cases} e^\xi & \text{at } \xi < 0 \text{ and } |\xi| \gg 1, \\ \xi & \text{at } \xi > 0 \text{ and } |\xi| \gg 1. \end{cases} \quad \text{(A.6b)}$$

Using Eqs.(A.6), one could get from Eq.(A.5a) the following expression for the electrons concentration, valid under the condition $\left|E_F - D_0 - D_s - \frac{\Delta_s}{2}\right| \gg k_B T$:

$$n(E_F) \cong \sum_s \frac{1}{2\pi \hbar^2 v_s^2} \left[\Delta_s \left(E_F - D_0 - D_s - \frac{\Delta_s}{2}\right) + \left(E_F - D_0 - D_s - \frac{\Delta_s}{2}\right)^2\right] \equiv$$

$$\sum_s \frac{(E_F - D_0 - D_s)^2 - \left(\frac{\Delta_s}{2}\right)^2}{2\pi \hbar^2 v_s^2} \quad \text{at } E_F > D_0 + D_s + \frac{\Delta_s}{2} \quad \text{(A.7a)}$$

$$n(E_F) \cong \sum_s \frac{(k_B T)^2}{\pi \hbar^2 v_s^2} \left[\frac{\Delta_s}{2k_B T} + 1\right] e^{\frac{E_F - D_0 - D_s - \frac{\Delta_s}{2}}{k_B T}} \quad \text{at} \quad E_F < D_0 + D_s + \frac{\Delta_s}{2} \quad \text{(A.7b)}$$

In a similar way one could get from Eqs.(A.5b) and (A.6) the following expressions for the holes concentration, valid under the condition $\left|-E_F + D_0 + D_s - \frac{\Delta_s}{2}\right| \gg k_B T$:

$$p(E_F) \cong \sum_s \frac{1}{2\pi \hbar^2 v_s^2} \left[\Delta_s \left(-E_F + D_0 + D_s - \frac{\Delta_s}{2}\right) + \left(-E_F + D_0 + D_s - \frac{\Delta_s}{2}\right)^2\right] \equiv$$

$$\sum_s \frac{(-E_F + D_0 + D_s)^2 - \left(\frac{\Delta_s}{2}\right)^2}{2\pi \hbar^2 v_s^2} \quad \text{at } E_F < D_0 + D_s - \frac{\Delta_s}{2} \quad \text{(A.8a)}$$

$$p(E_F) \cong \sum_s \frac{(k_B T)^2}{\pi \hbar^2 v_s^2} \left[\frac{\Delta_s}{2k_B T} + 1\right] e^{\frac{-E_F + D_0 + D_s - \frac{\Delta_s}{2}}{k_B T}} \quad \text{at } E_F > D_0 + D_s - \frac{\Delta_s}{2} \quad \text{(A.8b)}$$

Comparison of Eqs.(A.7a) and (A.8b) shows that at $E_F > D_0 + D_s + \frac{\Delta_s}{2}$ the strong inequality is valid

$$n(E_F) \gg p(E_F) \text{ at } E_F > D_0 + D_s + \frac{\Delta_s}{2}, \quad \text{(A.9a)}$$

while at $E_F < D_0 + D_s - \frac{\Delta_s}{2}$ the situation is reverse

$$n(E_F) \ll p(E_F) \text{ at } E_F < D_0 + D_s - \frac{\Delta_s}{2}. \quad \text{(A.9b)}$$

Since the determination of Fermi energy $E_F$ needs the calculation of total charge density, Eqs.(A.7)-(A.9) allows one to write down the charge density $\rho(E_F) = e(p(E_F) - p(E_F))$:

$$\rho(E_F) \cong \begin{cases} \frac{e}{2\pi \hbar^2} \sum_s \frac{(E_F - D_0 - D_s)^2 - \left(\frac{\Delta_s}{2}\right)^2}{v_s^2} \text{sign}(D_0 + D_s - E_F), & |E_F - D_0 - D_s| > \frac{\Delta_s}{2} \\ \sum_s \frac{e(k_B T)^2}{\pi \hbar^2 v_s^2} \left[\frac{\Delta_s}{k_B T} + 2\right] e^{-\frac{\Delta_s}{2 k_B T}} \sinh\left(\frac{D_0 + D_s - E_F}{k_B T}\right), & |E_F - D_0 - D_s| \ll \frac{\Delta_s}{2} \end{cases} \quad \text{(A.10)}$$

Assuming that $v_s \approx v_F$, the approximate expression for Fermi energy should be found from the equation $\rho(E_F) + \sigma = 0$, that is a quadratic equation:

$$E_F^2 - E_F(2D_0 + D_\uparrow + D_\downarrow) + \frac{1}{2}\left[(D_0 + D_\uparrow)^2 + (D_0 + D_\downarrow)^2 - \left(\frac{\Delta_\uparrow}{2}\right)^2 - \left(\frac{\Delta_\downarrow}{2}\right)^2\right] - \frac{\pi \hbar^2 v_F^2 \sigma}{e} \cong 0 \quad \text{(A.11)}$$

The solution of Eq.(A.11) is

$$E_F(\sigma) \approx D_0 + \frac{D_\uparrow + D_\downarrow}{2} +$$

$$\sqrt{\left(D_0 + \frac{D_\uparrow + D_\downarrow}{2}\right)^2 + \frac{\pi \hbar^2 v_F^2 \sigma}{e} - \frac{1}{2}\left[(D_0 + D_\uparrow)^2 + (D_0 + D_\downarrow)^2 - \left(\frac{\Delta_\uparrow}{2}\right)^2 - \left(\frac{\Delta_\downarrow}{2}\right)^2\right]} = D_0 + \frac{D_\uparrow + D_\downarrow}{2} +$$

$$\sqrt{\frac{1}{8}[\Delta_\uparrow^2 + \Delta_\downarrow^2] - \left(\frac{D_\uparrow - D_\downarrow}{2}\right)^2 + \frac{\pi \hbar^2 v_F^2 \sigma}{e}} \quad \text{(A.12)}$$

It is seen from **Fig. A1** that approximate dependence for the charge density are rather close to exact ones.

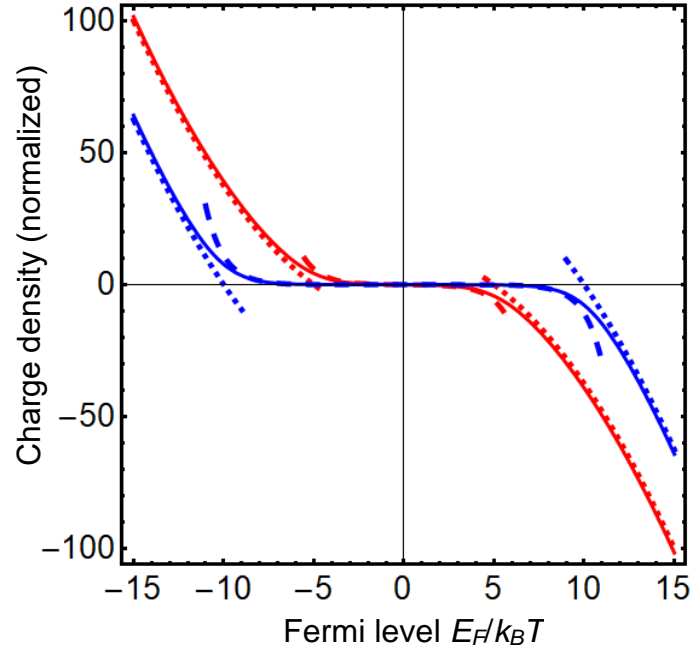

**Figure A1.** Total charge density as a function of normalized Fermi level for $\frac{\Delta_s}{k_B T}$=10 and 20 (red and blue curves respectively). Solid curves are plotted with exact expressions (A.5), while dotted and dashed curves are based on the expressions (7a), (8a) and (7b), (8b) respectively.